\documentclass[]{interact}
\pdfoutput=1
\usepackage{epstopdf}\usepackage[caption=false]{subfig}
\usepackage{natbib}
\bibpunct[, ]{(}{)}{;}{a}{}{,}
\usepackage{algorithm}
\usepackage[noend]{algpseudocode}
\usepackage{multirow}
\usepackage{url}

\theoremstyle{plain}

\theoremstyle{definition}

\theoremstyle{remark}

\begin{document}

\pdfinfo{
   /Author (Shane A. McQuarrie, Cheng Huang, Karen E. Willcox)
   /Title  (Data-driven reduced-order models via regularized operator inference for a single-injector combustion process)
   /Keywords (nonlinear model reduction, data-driven reduced model, scientific machine learning, operator inference, combustion, reacting flows)
}

\title{Data-driven reduced-order models via regularized operator inference for a single-injector combustion process}

\author{
\name{Shane~A.~McQuarrie\textsuperscript{a}, Cheng~Huang\textsuperscript{b}, Karen~E.~Willcox\textsuperscript{a}\thanks{CONTACT Karen~E.~Willcox. Email: kwillcox@oden.utexas.edu}}
\affil{\textsuperscript{a}Oden Institute for Computational Engineering and Sciences, University of Texas at Austin, USA; \textsuperscript{b}Department of Aerospace Engineering, University of Michigan, USA}
}

\maketitle

\begin{abstract}
This paper derives predictive reduced-order models for rocket engine combustion dynamics via Operator Inference, a scientific machine learning approach that blends data-driven learning with physics-based modeling. The non-intrusive nature of the approach enables variable transformations that expose system structure. The specific contribution of this paper is to advance the formulation robustness and algorithmic scalability of the Operator Inference approach. Regularization is introduced to the formulation to avoid over-fitting. The task of determining an optimal regularization is posed as an optimization problem that balances training error and stability of long-time integration dynamics. A scalable algorithm and open-source implementation are presented, then demonstrated for a single-injector rocket combustion example. This example exhibits rich dynamics that are difficult to capture with state-of-the-art reduced models. With appropriate regularization and an informed selection of learning variables, the reduced-order models exhibit high accuracy in re-predicting the training regime and acceptable accuracy in predicting future dynamics, while achieving close to a million times speedup in computational cost. When compared to a state-of-the-art model reduction method, the Operator Inference models provide the same or better accuracy at approximately one thousandth of the computational cost.
\end{abstract}


\section{Introduction} \label{sec:intro}
The emerging field of scientific machine learning brings together the perspectives of physics-based modeling and data-driven learning. In the field of fluid dynamics, physics-based modeling and simulation have played a critical role in advancing scientific discovery and driving engineering innovation in domains as diverse as  biomedical engineering \citep{yin2010simulation,nordsletten2011fluid}, geothermal modeling \citep{sullivan2001state,CFOS2011geothermalbayes}, and aerospace \citep{spalart2016role}. These advances are based on decades of mathematical and algorithmic developments in computational fluid dynamics (CFD).
Scientific machine learning builds upon these rigorous physics-based foundations while seeking to exploit the flexibility and expressive modeling capabilities of machine learning \citep{baker2019workshop}. This paper presents a scientific machine learning approach that blends data-driven learning with the theoretical foundations of physics-based model reduction. This creates the capability to learn predictive reduced-order models (ROMs) that provide approximate predictions of complex physical phenomena while exhibiting several orders of magnitude computational speedup over CFD.

Projection-based model reduction considers the class of problems for which the governing equations are known and for which we have a high-fidelity (e.g., CFD) model \citep{antoulas2005approximation, BGW2015pmorSurvery}. The goal is to derive a ROM that has lower complexity and yields accurate solutions with reduced computation time.
Projection-based approaches define a low-dimensional manifold on which the dominant dynamics evolve. This manifold may be defined as a function of the operators of the high-fidelity model, as in interpolatory methods that employ a Krylov subspace \citep{Bai02,Fre03}, or it may be determined empirically from representative high-fidelity simulation data, as in the proper orthogonal decomposition (POD) \citep{lumley,sirovich1987turbulence,berkooz}. The POD has been particularly successful in fluid dynamics, dating back to the early applications in unsteady flows and turbulence modeling \citep{sirovich1987turbulence,deane1991low,gatski1992pod-turbulence}, and in unsteady fluid-structure interaction \citep{dowell2001modeling}.

Model reduction methods have advanced to include error estimation \citep{Veroy03,Veroy05,Grepl05,RozHP08} and to address parametric and nonlinear problems \citep{BMNP2004eim,Astrid2008,CS2010deim,Carlberg2013}, yet the intrusive nature of the methods has limited their impact in practical applications. When legacy or commercial codes are used, as is often the case for CFD applications, it can be difficult or impossible to implement classical projection-based model reduction. Black-box surrogate modeling instead derives the ROM by fitting to simulation data; such methods include response surfaces and Gaussian process models, long used in engineering, as well as machine learning surrogate models. These methods are powerful and often yield good results, but since the approximations are based on generic data-fit representations, they are not equipped with the guarantees (e.g., stability guarantees, error estimators) that accompany projection-based ROMs. Nonlinear system identification techniques seek to illuminate the black box by discovering the underlying physics of a system from data \citep{BPK2016sindy}. However, when the governing dynamics are known and simulation data are available, reduced models may be directly tailored to the specific dynamics without access to the details of the large-scale CFD code.

This paper presents a non-intrusive alternative to black-box surrogate modeling. We use the Operator Inference method of \cite{PW2016operatorInference} to learn a ROM from simulation data; the structure of the ROM is defined by known governing equations combined with the theory of projection-based model reduction. Our approach may be termed `glass-box modeling', which we define as the situation where the form of the targeted dynamics is known (here via the partial differential equations that define the problem of interest), but we do not have internal access to the CFD code that produces the simulation data. That is, we know what dynamics to expect, but we may only calibrate our models using outputs of the full-order CFD model. This glass-box setting is in contrast to black-box modeling approaches which do not exploit knowledge of the governing equations. We build on our prior work in \cite{SKHW2020romCombustion} by formally introducing regularization to the Operator Inference approach.  Regularization is critical to avoid overfitting for problems with complex dynamics, as is the case for the combustion example considered here. A second contribution of this paper is a scalable implementation of the approach, which is available via an open-source implementation.
Section~\ref{sec:method} presents the methodology and regularization approach and describes the scalable implementation. Section~\ref{sec:results} presents numerical results for a single-injector combustion problem and Section~\ref{sec:conclusions} concludes the paper.

\section{Methodology} \label{sec:method}
This section begins with an overview of the Operator Inference approach in Section~\ref{subsec:operator-inference}. Section~\ref{subsec:regularization} augments Operator Inference with a new regularization formulation, posed as an optimization problem, and presents a complete algorithm for regularization selection and model learning. In Section~\ref{subsec:scalable-implementation}, we discuss a scalable implementation of the algorithm, which can then be applied to CFD problems of high dimension.

\subsection{Operator Inference}
\label{subsec:operator-inference}

We target problems governed by systems of nonlinear partial differential equations. Consider the governing equations of the system of interest written, after spatial discretization, in semi-discrete form
\begin{align}
    \label{eq:poly_FOM}
    \frac{\text{d}}{\text{d}t}\mathbf{q}(t)
    &= \mathbf{c}
     + \mathbf{A}\mathbf{q}(t)
     + \mathbf{H}(\mathbf{q}(t) \otimes \mathbf{q}(t))
     + \mathbf{B}\mathbf{u}(t),
    &
    \mathbf{q}(t_0)
    &= \mathbf{q}_0,
    &
    t &\in [t_0,t_f],
\end{align}
where
$\mathbf{q}(t) \in \mathbb{R}^{n}$ is the state vector discretized over $n$ spatial points at time $t$,
$\mathbf{u}(t) \in \mathbb{R}^m$ denotes the $m$ inputs at time $t$, typically related to boundary conditions or forcing terms,
$t_0$ and $t_f$ are respectively the initial and final time,
and $\mathbf{q}_0$ is the given initial condition.
We refer to Eq.~(\ref{eq:poly_FOM}) as the full-order model (FOM) and note that it has been written to have a polynomial structure:
$\mathbf{c}\in \mathbb{R}^{n}$ are constant terms;
$\mathbf{A} \mathbf{q}(t)$ are the terms that are linear in the state $\mathbf{q}(t)$, with the discretized operator $\mathbf{A}\in \mathbb{R}^{n \times n}$;
$\mathbf{H} (\mathbf{q}(t)\otimes \mathbf{q}(t))$ are the terms that are quadratic in $\mathbf{q}(t)$, with $\mathbf{H} \in \mathbb{R}^{n \times n^2}$;
and $\mathbf{B}\mathbf{u}(t)$ are the terms that are linear in the input $\mathbf{u}(t)$, with $\mathbf{B}\in \mathbb{R}^{n \times m}$.
This polynomial structure arises in three ways: (1)~it may be an attribute of the governing equations; (2)~it may be exposed via variable transformations; or (3)~it may be derived by introducing auxiliary variables through the process of lifting. As examples of each: (1)~the incompressible Navier-Stokes equations have a quadratic form; (2)~the Euler equations can be transformed to quadratic form by using pressure, velocity, and specific volume as state variables; (3)~a nonlinear tubular reactor model with Arrhenius reaction terms can be written in quadratic form via the lifting transformation shown in \cite{Kramer-AIAA2019} that introduces six auxiliary variables. Higher-order terms may be included in the formulation as well, but in this work we focus on a quadratic model due to the nature of the driving application.

A projection-based reduced-order model (ROM) of Eq.~\eqref{eq:poly_FOM} preserves the polynomial structure \citep{BGW2015pmorSurvery}. Approximating the high-dimensional state $\mathbf{q}$ in a low-dimensional basis $\mathbf{V}\in \mathbb{R}^{n \times r}$, with $r\ll n$, we write $\mathbf{q}(t) \approx \mathbf{V}\mathbf{q}_r(t)$, where $\mathbf{q}_r(t) \in \mathbb{R}^{r}$ is the reduced state. Using a Galerkin projection, this yields the \emph{intrusive} ROM of Eq.~\eqref{eq:poly_FOM}:
\begin{align*}
    \frac{\text{d}}{\text{d}t}\mathbf{q}_r(t)
    &= \mathbf{c}_r
     + \mathbf{A}_r \mathbf{q}_r(t)
     + \mathbf{H}_r (\mathbf{q}_r(t)\otimes \mathbf{q}_r(t))
     + \mathbf{B}_r\mathbf{u}(t),
    &
    \mathbf{q}_r(t_0)
    &= \mathbf{V}^\top\mathbf{q}_0,
    &
    t &\in [t_0,t_f],
\end{align*}
where
$\mathbf{c}_r = \mathbf{V}^\top \mathbf{c} \in \mathbb{R}^r$,
$\mathbf{A}_r = \mathbf{V}^\top \mathbf{A} \mathbf{V} \in \mathbb{R}^{r \times r}$,
$\mathbf{H}_r = \mathbf{V}^\top \mathbf{H} (\mathbf{V} \otimes \mathbf{V}) \in \mathbb{R}^{r \times r^2}$,
and $\mathbf{B}_r = \mathbf{V}^\top \mathbf{B} \in \mathbb{R}^{r\times m}$
are the ROM operators corresponding to the FOM operators $\mathbf{c}$, $\mathbf{A}$, $\mathbf{H}$, and $\mathbf{B}$, respectively. The ROM is intrusive because computing these ROM operators requires access to the discretized FOM operators, which typically entails intrusive queries and/or access to source code.

The \emph{non-intrusive} Operator Inference (OpInf) approach proposed by \cite{PW2016operatorInference} parallels the intrusive projection-based ROM setting, but learns ROMs from simulation data without direct access to the FOM operators. Recognizing that the intrusive ROM has the same polynomial form as Eq.~\eqref{eq:poly_FOM}, OpInf uses a data-driven regression approach to derive a ROM of Eq.~\eqref{eq:poly_FOM} as
\begin{align}
    \label{eq:poly_ROM}
    \frac{\text{d}}{\text{d}t}\widehat{\mathbf{q}}(t)
    &= \widehat{\mathbf{c}}
     + \widehat{\mathbf{A}} \widehat{\mathbf{q}}(t)
     + \widehat{\mathbf{H}} (\widehat{\mathbf{q}}(t)\otimes \widehat{\mathbf{q}}(t))
     + \widehat{\mathbf{B}}\mathbf{u}(t),
    &
    \widehat{\mathbf{q}}(t_0)
    &= \mathbf{V}^\top\mathbf{q}_0,
    &
    t &\in [t_0,t_f],
\end{align}
where
$\widehat{\mathbf{c}}\in\mathbb{R}^r$,
$\widehat{\mathbf{A}}\in\mathbb{R}^{r\times r}$,
$\widehat{\mathbf{H}}\in\mathbb{R}^{r\times \binom{r+1}{2}}$,
and $\widehat{\mathbf{B}}\in\mathbb{R}^{r\times m}$
are determined by solving a data-driven regression problem, and $\widehat{\mathbf{q}}(t) \in \mathbb{R}^{r}$ is the state of the OpInf ROM.\footnote{From here on we use $\widehat{\mathbf{q}}\otimes\widehat{\mathbf{q}}$ to indicate a compact Kronecker product with only the $\binom{r+1}{2}=\frac{1}{2}r(r+1)$ unique quadratic terms ($\hat{q}_1^2$, $\hat{q}_1 \hat{q}_2$, $\hat{q}_1 \hat{q}_3$, $\ldots$); for matrices, the product is applied column-wise.}

OpInf solves a regression problem to find reduced operators that yield the ROM that best matches projected snapshot data in a minimum-residual sense. Mathematically, OpInf solves the least-squares problem
\begin{align}
    \label{eq:OpInf}
    \min_{ \widehat{\mathbf{c}}, \widehat{\mathbf{A}}, \widehat{\mathbf{H}}, \widehat{\mathbf{B}} }\sum_{j=0}^{k-1}\left\|
      \widehat{\mathbf{c}}
    + \widehat{\mathbf{A}}\widehat{\mathbf{q}}_j
    + \widehat{\mathbf{H}}(\widehat{\mathbf{q}}_j\otimes\widehat{\mathbf{q}}_j)
    + \widehat{\mathbf{B}}\mathbf{u}_j
    - \dot{\widehat{\mathbf{q}}}_j
    \right\|_2^2,
\end{align}
where
$\{\widehat{\mathbf{q}}_j\}_{j=0}^{k-1}$ is the dataset used to drive the learning with $\widehat{\mathbf{q}}_j$ denoting a reduced-state snapshot at timestep $j$,
$\{\dot{\widehat{\mathbf{q}}}_j\}_{j=0}^{k-1}$ are the associated time derivatives,
and $\{\mathbf{u}_j\}_{j=0}^{k-1}$ is the collection of inputs corresponding to the data with $\mathbf{u}_j \equiv \mathbf{u}(t_j)$.
To generate the dataset, we employ the following steps:
(1)~Collect a set of $k$ high-fidelity state snapshots $\{\mathbf{z}_{j}\}_{j=0}^{k-1}\subset\mathbb{R}^{\tilde{n}}$ by solving the original high-fidelity model at times $\{t_j\}_{j=0}^{k-1}$ with inputs $\{\mathbf{u}_j\}_{j=0}^{k-1}$.
(2)~Apply a variable transformation $\mathbf{q}_{j} = \mathcal{T}(\mathbf{z}_j)$ to obtain snapshots of the transformed variables. Here $\mathcal{T}:\mathbb{R}^{\tilde{n}}\to\mathbb{R}^{n}$ is the map representing a reversible transformation (e.g., from density to specific volume) or a lifting transformation \citep{QKPW2020liftAndLearn}.
(3)~Compute the proper orthogonal decomposition (POD) basis $\mathbf{V}$ of the transformed snapshots.\footnote{Or any other low-dimensional basis as desired.}
(4)~Project the transformed snapshots onto the POD subspace as $\widehat{\mathbf{q}}_j = \mathbf{V}^\top\mathbf{q}_j$. (5)~Estimate projected time derivative information $\{\dot{\widehat{\mathbf{q}}}\}_{j=0}^{k-1}$.
The training period $[t_0,t_{k-1}]$ for which we have data is a subset of the full time domain of interest $[t_0,t_f]$; results from the ROM over $[t_k,t_f]$ will be entirely predictive.

Eq.~\eqref{eq:OpInf} can also be written in matrix form as
\begin{align}
    \label{eq:OpInf-matrix}
    \min_{\mathbf{O}}
    \left\|
      \mathbf{D}\mathbf{O}^\top
      - \mathbf{R}^\top
    \right\|_F^2,
\end{align}
where
\begin{align*}
    \mathbf{O}
    &= \left[\begin{array}{cccc}
        \widehat{\mathbf{c}} &
        \widehat{\mathbf{A}} &
        \widehat{\mathbf{H}} &
        \widehat{\mathbf{B}}
    \end{array}\right]\in\mathbb{R}^{r\times d(r,m)},
    &\text{(unknown operators)}
    \\
    \mathbf{D}
    &= \left[\begin{array}{cccc}
        \mathbf{1}_k &
        \widehat{\mathbf{Q}}^\top &
        (\widehat{\mathbf{Q}}\otimes\widehat{\mathbf{Q}})^\top &
        \mathbf{U}^\top
    \end{array}\right]\in\mathbb{R}^{k\times d(r,m)},
    &\text{(known data)}
    \\
    \widehat{\mathbf{Q}}
    &= \left[\begin{array}{cccc}
        \widehat{\mathbf{q}}_0 & \widehat{\mathbf{q}}_1 & \cdots & \widehat{\mathbf{q}}_{k-1}
    \end{array}\right]\in\mathbb{R}^{r\times k},
    &\text{(snapshots)}
    \\
    \mathbf{R}
    &= \left[\begin{array}{cccc}
        \dot{\widehat{\mathbf{q}}}_0 & \dot{\widehat{\mathbf{q}}}_1 & \cdots & \dot{\widehat{\mathbf{q}}}_{k-1}
    \end{array}\right]\in\mathbb{R}^{r\times k},
    &\text{(time derivatives)}
    \\
    \mathbf{U}
    &= \left[\begin{array}{cccc}
        \mathbf{u}_0 & \mathbf{u}_1 & \cdots & \mathbf{u}_{k-1}
    \end{array}\right]\in\mathbb{R}^{m\times k},
    &\text{(inputs)}
\end{align*}
and where $d(r,m) = 1 + r + \binom{r+1}{2} + m$ and $\mathbf{1}_k \in \mathbb{R}^{k}$ is the length-$k$ column vector with all entries set to unity. The OpInf least-squares problem Eq.~\eqref{eq:OpInf} is therefore linear in the coefficients of the unknown ROM operators $\widehat{\mathbf{c}}$, $\widehat{\mathbf{A}}$, $\widehat{\mathbf{H}}$, and $\widehat{\mathbf{B}}$.

The OpInf approach permits us to compute the ROM operators  $\widehat{\mathbf{c}}$, $\widehat{\mathbf{A}}$, $\widehat{\mathbf{H}}$, and $\widehat{\mathbf{B}}$ without explicit access to the original high-dimensional operators $\mathbf{c}$, $\mathbf{A}$, $\mathbf{H}$, and $\mathbf{B}$. This point is key since we apply variable transformations only to the snapshot data, not to the operators or the underlying model. Thus, even in a setting where deriving a classical intrusive ROM might be possible, the OpInf approach enables us to work with variables other than those used for the original high-fidelity discretization. In Section~\ref{sec:results} we will see the importance of this for a reacting flow application. We also note that under some conditions, OpInf recovers the intrusive POD ROM \citep{PW2016operatorInference,Peherstorfer2019reprojection}.

\subsection{Regularization}
\label{subsec:regularization}

The problem in Eq.~\eqref{eq:OpInf-matrix}
is generally overdetermined (i.e., $k > d(r,m)$), but is also noisy due to errors in the numerically estimated time derivatives $\mathbf{R}$, model mis-specification (e.g., if the system is not truly quadratic), and truncated POD modes that leave some system dynamics unresolved.
The ROMs resulting from Eq.~\eqref{eq:OpInf-matrix} can thus suffer from overfitting the operators to the data and therefore exhibit poor predictive performance over the time domain of interest $[t_0,t_f]$.

To avoid overfitting, we introduce a Tikhonov regularization \citep{Tikhonov1977regularization} to
Eq.~\eqref{eq:OpInf-matrix}, which then becomes
\begin{align}
    \label{eq:OpInf-reg}
\min_{\mathbf{O}}
    \left\|\mathbf{D}\mathbf{O}^\top - \mathbf{R}^\top\right\|_{F}^2
    + \left\|\boldsymbol{\Gamma}\mathbf{O}^\top\right\|_{F}^2,
\end{align}
where
$\boldsymbol{\Gamma} \in \mathbb{R}^{d(r,m)\times d(r,m)}$ is a full-rank regularizer.
The minimizer of Eq.~\eqref{eq:OpInf-reg} satisfies the modified normal equations
\begin{align}
    \label{eq:normal-equations}
    \left(\mathbf{D}^\top\mathbf{D} + \boldsymbol{\Gamma}^\top\boldsymbol{\Gamma}\right)\mathbf{O}^\top
    = \mathbf{D}^\top\mathbf{R}^\top,
\end{align}
which admit a unique solution since $\mathbf{D}^\top\mathbf{D} + \boldsymbol{\Gamma}^\top\boldsymbol{\Gamma}$ is symmetric positive definite.

An $L_2$ regularizer $\boldsymbol{\Gamma} = \lambda \mathbf{I}$, $\lambda > 0$ and $\mathbf{I}$ the identity matrix, penalizes each entry of the inferred ROM operators $\widehat{\mathbf{c}}$, $\widehat{\mathbf{A}}$, $\widehat{\mathbf{H}}$, and $\widehat{\mathbf{B}}$, thereby driving the ROM toward the globally stable system $\frac{\textrm{d}}{\textrm{d}t}\widehat{\mathbf{q}}(t) = \mathbf{0}$.
Since the entries of the quadratic operator $\widehat{\mathbf{H}}$ have a different scaling than entries of the other operators, we construct a diagonal regularizer $\boldsymbol{\Lambda}(\lambda_1,\lambda_2)$, with $\lambda_1,\lambda_2 > 0$, such that the operator entries are penalized by $\lambda_1$, except for the entries of $\widehat{\mathbf{H}}$, which are penalized by $\lambda_2$. That is, with $\boldsymbol{\Gamma} = \boldsymbol{\Lambda}(\lambda_1,\lambda_2)$, Eq.~\eqref{eq:OpInf-reg} can be expressed as
\begin{align*}
    \min_{
        \widehat{\mathbf{c}},
        \widehat{\mathbf{A}},
        \widehat{\mathbf{H}},
        \widehat{\mathbf{B}}
        }
    \left[\left\|\mathbf{D}
    \left[\begin{array}{cccc}
          \widehat{\mathbf{c}}
        & \widehat{\mathbf{A}}
        & \widehat{\mathbf{H}}
        & \widehat{\mathbf{B}}
    \end{array}\right]^\top
    - \mathbf{R}^\top
    \right\|_{F}^2
    + \lambda_{1}\left(\|\widehat{\mathbf{c}}\|_{2}^{2}
    + \|\widehat{\mathbf{A}}\|_{F}^{2}
    + \|\widehat{\mathbf{B}}\|_{F}^{2}\right)
    + \lambda_{2}\|\widehat{\mathbf{H}}\|_{F}^{2}\right].
\end{align*}

The scalar hyperparameters $\lambda_{1}$ and $\lambda_{2}$, which balance the minimization between the data fit and the regularization, must be chosen with care.
The ideal regularizer produces a ROM that minimizes some error metric over the full time domain $[t_0,t_f]$; however, since data are only available for the smaller training domain $[t_0,t_{k-1}]$, we choose $\lambda_{1}$ and $\lambda_{2}$ so that the resulting ROM minimizes error over $[t_0,t_{k-1}]$ while maintaining a bound on the integrated POD coefficients over $[t_0,t_f]$.
That is, we require the state
$
    \widehat{\mathbf{q}}(t)
    = \left[\begin{array}{cccc}
        \hat{q}_1(t) & \hat{q}_2(t) & \cdots & \hat{q}_r(t)
    \end{array}\right]^\top
$
produced by integrating Eq.~\eqref{eq:poly_ROM} to satisfy $|\hat{q}_i(t)|\le B$, $i=1,\ldots,r$ and $t\in[t_0,t_f]$, for some $B > 0$.
This in turn ensures a bound on the magnitude of the entries of the high-dimensional state $\mathbf{q}(t) = \mathbf{V}\widehat{\mathbf{q}}(t)$:
\begin{align}
    \label{eq:state-bounding}
    |q_i(t)|
    &= \left|\sum_{j=1}^{r}V_{ij}\hat{q}_j(t)\right|
    \le \sum_{j=1}^{r}\left|V_{ij}\right|\left|\hat{q}_j(t)\right|
    \le B\sum_{j=1}^{r}\left|V_{ij}\right|,
    \qquad
    i = 1, 2, \ldots, n,
\end{align}
where $V_{ij}$ is the $i$th element of the $j$th POD basis vector.
Note that the bound $B$ may be chosen with the intent of imposing a particular bound on the $|q_i|$ since the sums $\sum_{j=1}^{r}|V_{ij}|$ can be precomputed. For example, our regularization strategy provides a computationally efficient way to impose the temperature limiters proposed by \cite{HDM2019poddeim-robustness}.

Algorithm \ref{alg:OpInf-reg} details our regularized OpInf procedure, in which we choose $B$ as a multiple of the maximum absolute entry of the projected training data $\widehat{\mathbf{Q}}$. This particular strategy for selecting $\lambda_{1}$ and $\lambda_{2}$ could be replaced with a cross-validation or resampling grid search technique, but our experiments in this vein did not produce robust results. The training error $\|\widehat{\mathbf{Q}} - \widetilde{\mathbf{Q}}_{:,:k}\|$ in step \ref{step:error-norm} may compare $\widehat{\mathbf{Q}}$ and $\widetilde{\mathbf{Q}}_{:,:k}$ directly in the reduced space (e.g., with a matrix norm or an $L^p([t_0,t_{k-1}])$ norm), or it may be replaced with a more targeted comparison of some quantity of interest.

\begin{algorithm}
\begin{algorithmic}[1]
\Procedure{RegOpInf}{snapshots $\mathbf{Z}\in\mathbb{R}^{\tilde{n}\times k}$,
    inputs $\mathbf{U}\in\mathbb{R}^{m\times k}$,
    final time $t_f > t_0$,
    reversible map $\mathcal{T}:\mathbb{R}^{\tilde{n}}\to\mathbb{R}^{n}$,
    reduced dimension $r\in\mathbb{N}$,
    bound margin $\tau \ge 1$
    }
    \State $\mathbf{Q} \gets \mathcal{T}(\mathbf{Z})$
        \label{step:transform}
        \Comment{Map native variables to learning variables (columnwise).}
    \State $\mathbf{V} \gets\ $\texttt{pod(}$\mathbf{Q},r$\texttt{)}
        \label{step:pod-basis}
        \Comment{Compute a rank-$r$ POD basis from transformed snapshots.}
    \State $\widehat{\mathbf{Q}} \gets \mathbf{V}^\top\mathbf{Q}$
        \label{step:project}
        \Comment{Project snapshots to the $r$-dimensional POD subspace.}
    \State $\mathbf{R} \gets \frac{\textrm{d}}{\textrm{d}t}\widehat{\mathbf{Q}}$
        \label{step:derivatives}
        \Comment{Estimate time derivatives of the projected snapshots.}
\State $B \gets \tau\max_{i,j}|\widehat{\mathbf{Q}}_{ij}|$
        \Comment{Select a bound to require for integrated POD modes.}
        \label{step:select-bound}
\Procedure{TrainError}{$\lambda_1,\lambda_2$}
        \label{step:subroutine}
        \State $\widehat{\mathbf{c}}, \widehat{\mathbf{A}}, \widehat{\mathbf{H}}, \widehat{\mathbf{B}} \gets$ solve Eq.~\eqref{eq:OpInf-reg}
            with regularizer $\boldsymbol{\Gamma} = \boldsymbol{\Lambda}(\lambda_{1}, \lambda_{2})$
            \label{step:trainsolve}
        \State $\widetilde{\mathbf{Q}} \gets\ $ integrate Eq.~\eqref{eq:poly_ROM} with $\widehat{\mathbf{c}}, \widehat{\mathbf{A}}, \widehat{\mathbf{H}}, \widehat{\mathbf{B}}$ from $\widehat{\mathbf{q}}_0 = \widehat{\mathbf{Q}}_{:,0}$ over $[t_0,t_f]$
        \label{step:integrate}
        \If{$\max_{i,j}|\widetilde{\mathbf{Q}}_{ij}| > B$}
            \State \textbf{return} $\infty$
            \Comment{Disqualify ROMs that violate the bound.}
        \Else
            \State \textbf{return} $\|\widehat{\mathbf{Q}} - \widetilde{\mathbf{Q}}_{:,:k}\|$
                \label{step:error-norm}
                \Comment{Compute ROM error over $[t_0,t_{k-1}]$.}
        \EndIf
    \EndProcedure
    \State $\lambda_{1}^\ast,\lambda_{2}^\ast \gets \textrm{argmin}$ \textproc{TrainError}$(\lambda_{1},\lambda_{2})$
        \label{step:minimization}
    \State $\widehat{\mathbf{c}}, \widehat{\mathbf{A}}, \widehat{\mathbf{H}}, \widehat{\mathbf{B}} \gets$ solve Eq.~\eqref{eq:OpInf-reg}
        with optimal regularizer $\boldsymbol{\Gamma} = \boldsymbol{\Lambda}(\lambda_{1}^{\ast},\lambda_{2}^{\ast})$
    \label{step:finalsolve}
    \State \textbf{return} $\widehat{\mathbf{c}}, \widehat{\mathbf{A}}, \widehat{\mathbf{H}}, \widehat{\mathbf{B}}$
\EndProcedure
\end{algorithmic}
\caption{Operator Inference with regularization selection}
\label{alg:OpInf-reg}
\end{algorithm}

The regularization approach of Algorithm \ref{alg:OpInf-reg} may be viewed as a stabilization method since it selects a ROM with reasonable behavior over a given time domain. It should be noted, however, that this method does not modify an existing ROM to achieve stability, different from other stabilization methods such as eigenvalue reassignment \citep{KvBWAB2014eigreassign,rezaian2020hybrid}. The optimization is driven by a penalization but has no built-in constraints, which is a major advantage in terms of the computational cost, but the resulting ROMs are not guaranteed to preserve properties such as energy conservation. Adding constraints to Operator Inference to target conservation, similar to the work in \cite{CCS2018conservativeROM}, is a subject for possible future work.

\subsection{Scalable Implementation}
\label{subsec:scalable-implementation}

The steps of Algorithm \ref{alg:OpInf-reg} are highly modular and amenable to large-scale problems.
The variable transformation $\mathbf{Q} = \mathcal{T}(\mathbf{Z})$ of step \ref{step:transform} consists of $O(nk)$ elementary computations on the original snapshot matrix $\mathbf{Z}\in\mathbb{R}^{\tilde{n}\times k}$.
To compute the rank-$r$ POD basis $\mathbf{V}\in\mathbb{R}^{n\times r}$ in step \ref{step:pod-basis}, we use a randomized singular-value decomposition algorithm requiring $O(r n k + r^2(n + k))$ operations \citep{HMPT2011rNLA}; since $r\ll k,n$, the leading order behavior is $O(r n k)$.
The projection $\widehat{\mathbf{Q}} = \mathbf{V}^\top \mathbf{Q}$ in step \ref{step:project} costs about $2n k$ operations.
Note that $\widehat{\mathbf{Q}}\in\mathbb{R}^{r\times k}$ is small compared to $\mathbf{Q}\in\mathbb{R}^{n \times k}$, as typically $r\sim 10^0$--$10^3$ and $n\sim 10^4$--$10^9$.
The time derivatives $\mathbf{R}\in\mathbb{R}^{r\times k}$ in step \ref{step:derivatives} may be provided by the full-order solver or estimated, e.g., with finite differences of the columns of $\widehat{\mathbf{Q}}$.
In the latter case, the cost is $O(rk)$.
Step \ref{step:select-bound} is a simple $O(rk)$ selection.
The computational cost of steps \ref{step:transform}--\ref{step:select-bound} is therefore $O(rnk)$.

The subroutine \textproc{TrainError}$: \mathbb{R}^{2} \to \mathbb{R}$ comprising steps \ref{step:subroutine}--\ref{step:error-norm} must be evaluated for several choices of $\lambda_1$ and $\lambda_2$.
In step \ref{step:trainsolve}, we solve the minimization problem Eq.~\eqref{eq:OpInf-reg} via the linear system Eq.~\eqref{eq:normal-equations}.
Forming $\mathbf{D}^\top\mathbf{D} \in \mathbb{R}^{d(r,m)\times d(r,m)}$ and
$\mathbf{D}^\top\mathbf{R}^\top \in \mathbb{R}^{d(r,m) \times r}$ costs $O(d(r,m)^2 k + r d(r,m)k) = O(r^4 k)$ operations, but these can be precomputed once and reused in each subroutine evaluation; for specific $\lambda_1$ and $\lambda_2$, we form $\boldsymbol{\Gamma}\in\mathbb{R}^{d(r,m) \times d(r,m)}$ and compute $\mathbf{D}^\top\mathbf{D} + \boldsymbol{\Gamma}^\top \boldsymbol{\Gamma}\in\mathbb{R}^{d(r,m) \times d(r,m)}$ with only $2d(r,m)$ additional operations since $\boldsymbol{\Gamma}$ is diagonal.
Solving Eq.~\eqref{eq:normal-equations} via the Cholesky decomposition has a leading order cost of $\frac{1}{3}(d(r,m)^3) = \frac{1}{24}r^6$ operations \citep{demmel1997nla}, but this estimate can be improved upon with iterative methods for larger $r$ if desired.
The ROM integration in step \ref{step:integrate} can be carried out with any time-stepping scheme; for explicit methods, evaluating the ROM at a single point, i.e., computing the right-hand side of Eq.~\eqref{eq:poly_ROM}, costs $O(r^3)$ operations.
The total cost each evaluation of \textproc{TrainError}$: \mathbb{R}^{2} \to \mathbb{R}$ is therefore dominated by the cost of solving Eq.~\eqref{eq:normal-equations}, which is independent of $n$ and $k$.

Finally, the minimization in step \ref{step:minimization} is carried out with a derivative-free search method, which enables fewer total evaluations of the subroutine than a fine grid search. However, a coarse grid search is useful for identifying appropriate initial guesses for $\lambda_{1}$ and $\lambda_{2}$.

\section{Results} \label{sec:results}

\begin{figure}
\centering
    \includegraphics[width=\textwidth]{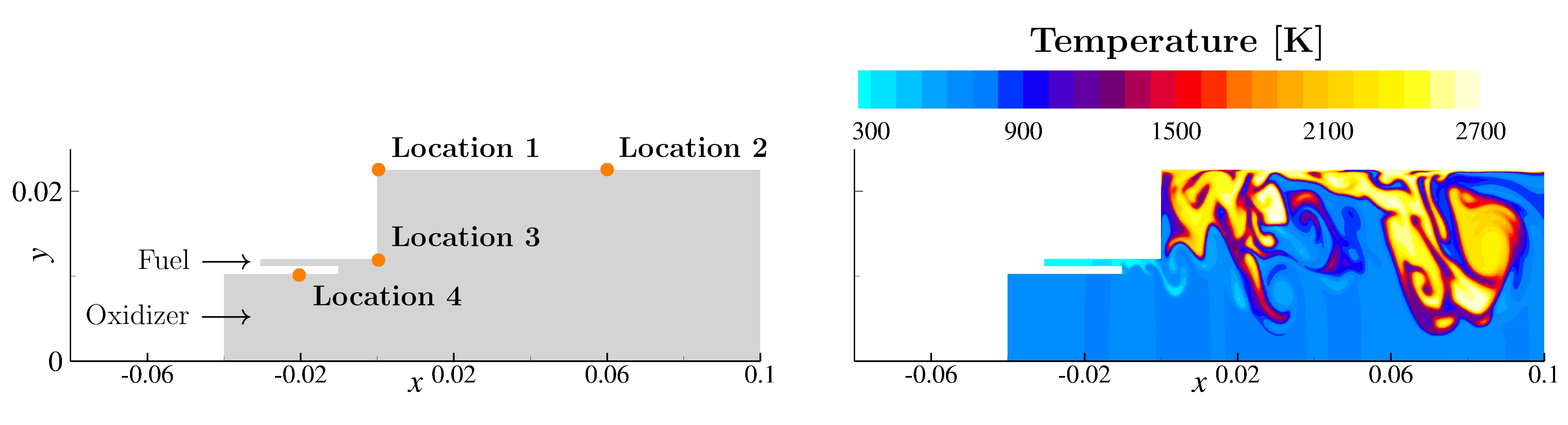}
\caption{The computational domain for the single-injector combustion problem. On the left, monitor locations for numerical results. On the right, a typical temperature field demonstrating the complexity and nonlinear nature of the problem.}
\label{fig:domain}
\end{figure}

This section applies regularized OpInf to a single-injector combustion problem, studied previously by \cite{SKHW2020romCombustion}, on the two-dimensional computational domain shown in Figure~\ref{fig:domain}.
Section \ref{subsec:problem-setup} describes the governing dynamics, a set of high-fidelity data obtained from a CFD code, and the variable transformations used to produce training data for learning reduced models with Algorithm \ref{alg:OpInf-reg}.
The resulting OpInf ROM performance is analyzed in Section~\ref{subsec:sensitivity-to-training-data} and compared to a state-of-the-art intrusive model reduction method in Section \ref{subsec:compare-poddeim}.

\subsection{Problem setup}
\label{subsec:problem-setup}

The combustion dynamics for this problem are governed by conservation laws
\begin{align}
    \label{eq:conservative}
    \frac{\partial\vec{q}_\textrm{c}}{\partial t}
    + \nabla \cdot(\vec{K} - \vec{K}_{\textrm{v}})
    &= \vec{S},
\end{align}
where
$
    \vec{q}_\textrm{c}
    = \left[\begin{array}{cccccccc}
        \rho & \rho v_x & \rho v_y & \rho e &
        \rho Y_1 & \rho Y_2 & \rho Y_3 & \rho Y_4
    \end{array}\right]^\top
$
are the conservative variables, $\vec{K}$ are the inviscid flux terms, $\vec{K}_\textrm{v}$ are the viscous flux  terms, and $\vec{S}$ are the source terms.
Here $\rho$ is the density $[\frac{\textrm{kg}}{\textrm{m}^3}]$,
$v_x$ and $v_y$ are the $x$ and $y$ velocity $[\frac{\textrm{m}}{\textrm{s}}]$,
$e$ is the total energy $[\frac{\textrm{J}}{\textrm{m}^3}]$,
and $Y_\ell$ is the $\ell$th species mass fraction with $\ell=1,2,3,4$.
The chemical species are CH$_4$, O$_2$, H$_2$O, and CO$_2$, which follow a global one-step chemical reaction
$
    \textrm{CH}_4 + 2\textrm{O}_2
    \to
    \textrm{CO}_2 + 2\textrm{H}_2\textrm{O}
$
\citep{WD1981oxidation}.
See \cite{HHSFAMT2015gems} for more details on the governing equations.

At the downstream end of the combustor, we impose a non-reflecting boundary condition while maintaining the chamber pressure via
\begin{align}
    p_\textrm{back}(t)
    &= p_{\textrm{back,ref}}\left(1 + 0.1\sin(2\pi f t)\right)
    \label{eq:input-function}
\end{align}
where $p_{\textrm{back,ref}} = 10^6$~Pa and $f = 5{,}000$~Hz.
The top and bottom wall boundary conditions are no-slip conditions, and for the upstream boundary we impose a constant mass flow at the inlets.

\cite{SKHW2020romCombustion} show that if the governing equations (\ref{eq:conservative}) are transformed to be written in the specific volume variables, many of the terms take a quadratic form.
Following that idea, we choose as learning variables the transformed and augmented state
$
\vec{q}= \left[\begin{array}{ccccccccc}
        p & v_x & v_y & T & \xi &
        c_1 & c_2 & c_3 & c_4
    \end{array}\right]^\top
$
where $p$ is the pressure $[\textrm{Pa}]$,
$T$ is the temperature $[\textrm{K}]$,
$\xi = 1/\rho$ is the specific volume $[\frac{\textrm{m}^3}{\textrm{kg}}]$,
and $c_1,\ldots,c_4$ are the species molar concentrations $[\frac{\textrm{kmol}}{\textrm{m}^3}]$ given by $c_\ell = \rho Y_\ell / M_\ell$ with $M_\ell$ the molar mass of the $\ell$th species $[\frac{\textrm{g}}{\textrm{mol}}]$.
As shown in \cite{SKHW2020romCombustion}, the equations for $v_x$, $v_y$, and $\xi$ are exactly quadratic in $\vec{q}$, while the remaining equations are quadratic with some non-polynomial terms in $\vec{q}$. Note that, differently from \cite{SKHW2020romCombustion}, $\vec{q}$ here is chosen to contain specific volume, pressure, and temperature, even though only two of the three quantities are needed to fully define the high-fidelity model (and the equation of state then defines the third). We augment the learning variables in this way because doing so exposes the quadratic form while also directly targeting the variables that are of primary interest for assessing ROM performance. In particular, the resulting ROMs provide more accurate predictions of temperature when temperature is included explicitly as a learning variable. We can do this since the transformations are applied only to the snapshot data, not to the CFD model itself. Indeed, constructing the full-order spatial operators in these transformed coordinates would be both impractical and inexact. This flexibility to learn from transformed snapshots instead of transformed full-order operators is a major advantage of the non-intrusive OpInf approach in comparison to traditional intrusive projection-based model reduction methods.

To generate high-fidelity training data, we use the finite-volume based General Equation and Mesh Solver (GEMS) \citep{HHSFAMT2015gems} to solve for the variables
$
\left[\begin{array}{cccccccc}
    p & v_x & v_y & T & Y_1 & Y_2 & Y_3 & Y_4
    \end{array}\right]$
over $n_x = 38{,}523$ cells, resulting in snapshots with $8n_x = 308{,}184$ entries each.
The snapshots are computed for 60,000 time steps beyond the initial condition with a temporal discretization of $\delta t = 10^{-7}$~s, from $t_0 = 0.015$~s to $t_f = 0.021$~s.
The computational cost of computing this dataset is approximately 1,200 CPU hours on two computing nodes, each of which contains two Haswell CPUs at $2.60$~GHz and $20$ cores per node.

Scaling is an essential aspect of successful model reduction and is particularly critical for this problem due to the wide range of scales across variables.
After transforming the GEMS snapshot data to the learning variables $\vec{q}$,
the species molar concentrations are scaled to $[0,1]$, and all other variables are scaled to $[-1,1]$.
This scaling ensures that null velocities and null molar concentrations are preserved.
For example, some upstream regions of the injector have zero methane concentration at all times.
By construction, the POD basis vectors and thus the ROM predictions will preserve those zero concentration values.

We implement Algorithm \ref{alg:OpInf-reg} via the \texttt{rom\_operator\_inference} Python package,\footnote{See \url{https://github.com/Willcox-Research-Group/rom-operator-inference-Python3}.} which is built on NumPy, SciPy, and scikit-learn \citep{WCV2011NumPy,2020SciPy-NMeth,sklearn-api}.
The time derivatives in step \ref{step:derivatives} of Algorithm \ref{alg:OpInf-reg} are estimated with fourth-order finite differences, and the least-squares problem in step \ref{step:trainsolve} are solved by applying the LAPACK routine POSV to Eq.~\eqref{eq:normal-equations}.
The learned ROMs are integrated in step \ref{step:integrate} with the explicit, adaptive, fourth-order Runge-Kutta scheme RK45, and the error evaluation of step \ref{step:error-norm} uses the $L^2([t_0,t_{k-1}])$ norm in the reduced space.
To minimize the function \textproc{TrainError}$:\mathbb{R}^{2}\to\mathbb{R}$ in step \ref{step:minimization}, we find an initial estimate of the minimizer via a coarse grid search, then refine the result with a Nelder-Mead search method \citep{nelder1965simplex}.
The code and details are publicly available at \url{https://github.com/Willcox-Research-Group/ROM-OpInf-Combustion-2D}.

\subsection{Sensitivity to Training Data}
\label{subsec:sensitivity-to-training-data}

We study the sensitivity of our approach to the training data by varying the number of snapshots used to compute the POD basis and learn the OpInf ROM. Specifically, we consider the three cases where we use the first $k = 10{,}000$, $k = 20{,}000$, and $k = 30{,}000$ snapshots from GEMS as training data sets. In each case we compute the POD basis and, to select an appropriate reduced dimension $r$, the cumulative energy based on the POD singular values:
$
    \mathcal{E}_{r}
    = \left.
    \left(\sum_{j=1}^{r}\sigma_j^2\right)
    \middle/
    \left(\sum_{j=1}^{k}\sigma_j^2\right)
    \right.
$, where $\{\sigma_{j}\}_{j=1}^{k}$ are the singular values of the learning variable snapshot matrix $\mathbf{Q}$ (see Figure~\ref{figure:singular-values}). Specifically, we choose the minimal integer $r$ such that $\mathcal{E}_{r}$ is greater than a fixed energy threshold. Table~\ref{table:cumulative-energy} shows that $r$ is increasing linearly with the number of snapshots, indicating that the basis is not being saturated as additional information is incorporated. This is an indication of the challenging nature of this application, due to the rich and complex dynamics. Table~\ref{table:cumulative-energy} also shows the column dimension $d(r,m)$ of the data matrix $\mathbf{D}\in\mathbb{R}^{k \times d(r,m)}$ in the Operator Inference problem, which grows quadratically with $r$; choosing $r$ so that $k \gg d(r,m)$ helps ensure that $\mathbf{D}$ has full rank.

\begin{table}
\begin{minipage}[b]{0.5\textwidth}
    \begin{figure}[H]
        \centering
        \includegraphics[width=\textwidth]{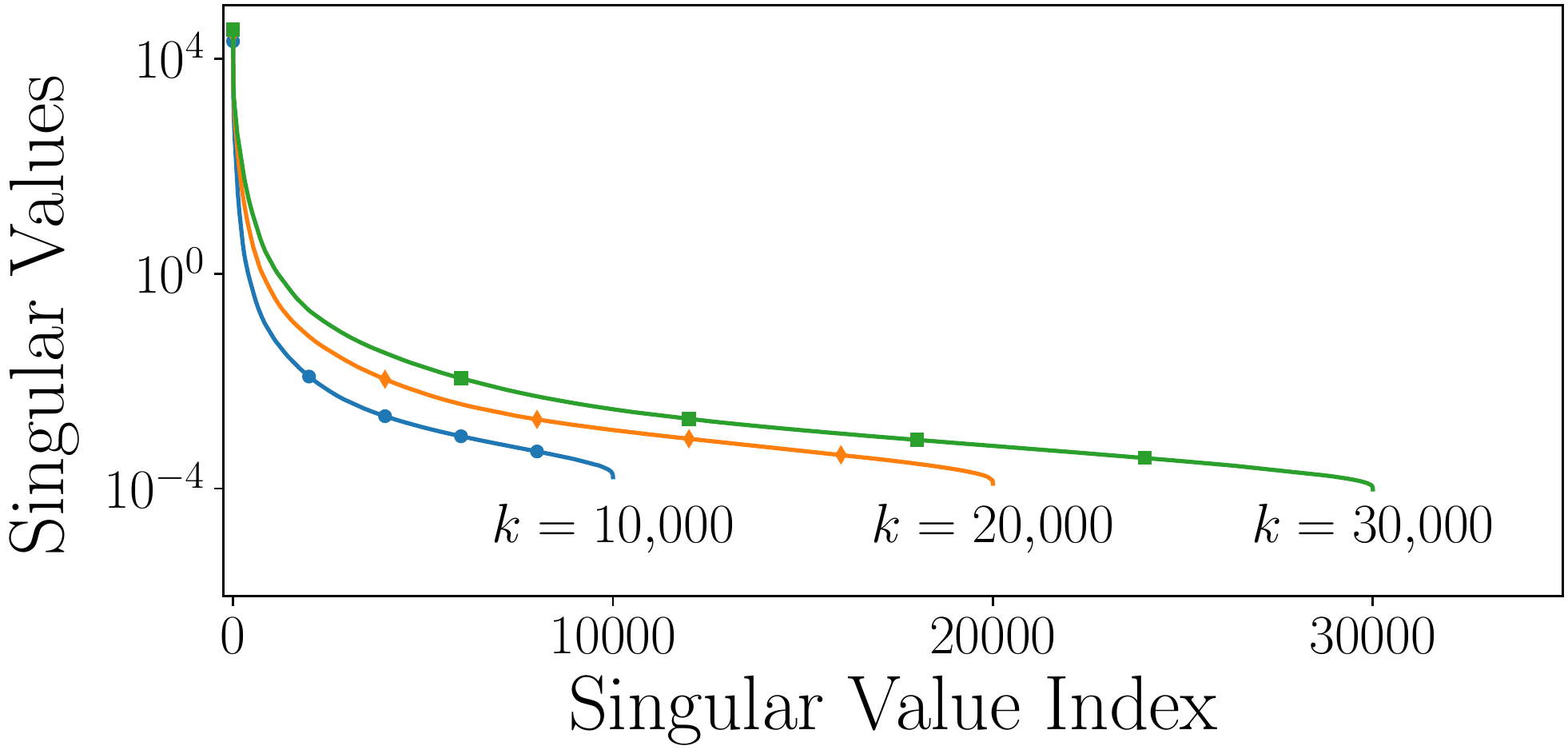}
        \vspace{-0.5cm}
        \caption{The POD singular values for varying size of the snapshot training set.}
        \label{figure:singular-values}
    \end{figure}
\end{minipage}
\hspace{.04\textwidth}
\begin{minipage}[b]{0.45\textwidth}
    \begin{table}[H]
    \centering
    \tbl{The basis size $r$ required to exceed a given cumulative energy level $\mathcal{E}_{r}$ increases linearly with the number of snapshots $k$ in the training set; the dimension $d = d(r,m)$ increases quadratically with $r$.\\}
    {\begin{tabular}{r||l|l|l} & \multicolumn{3}{c}{Cumulative energy $\mathcal{E}_{r}$} \\
            & \multicolumn{1}{c|}{$0.985$}
            & \multicolumn{1}{c|}{$0.990$}
            & \multicolumn{1}{c}{$0.995$}
        \\ \hline \hline
        \multirow{2}{*}{$k = 10{,}000$}
            & $r =  22$ & $r =  27$ & $r =  36$ \\
            & $d = 277$ & $d = 407$ & $d = 704$
        \\ \hline
        \multirow{2}{*}{$k = 20{,}000$}
            & $r =  43$ & $r =      53$ & $r =      72$ \\
            & $d = 991$ & $d = 1{,}486$ & $d = 2{,}701$
        \\ \hline
        \multirow{2}{*}{$k = 30{,}000$}
            & $r =      66$ & $r =      82$ & $r =     110$ \\
            & $d = 2{,}279$ & $d = 3{,}487$ & $d = 6{,}217$
\\
    \end{tabular}}
    \label{table:cumulative-energy}
    \end{table}
\end{minipage}
\end{table}

\begin{figure} \centering
    \includegraphics[width=1\textwidth]{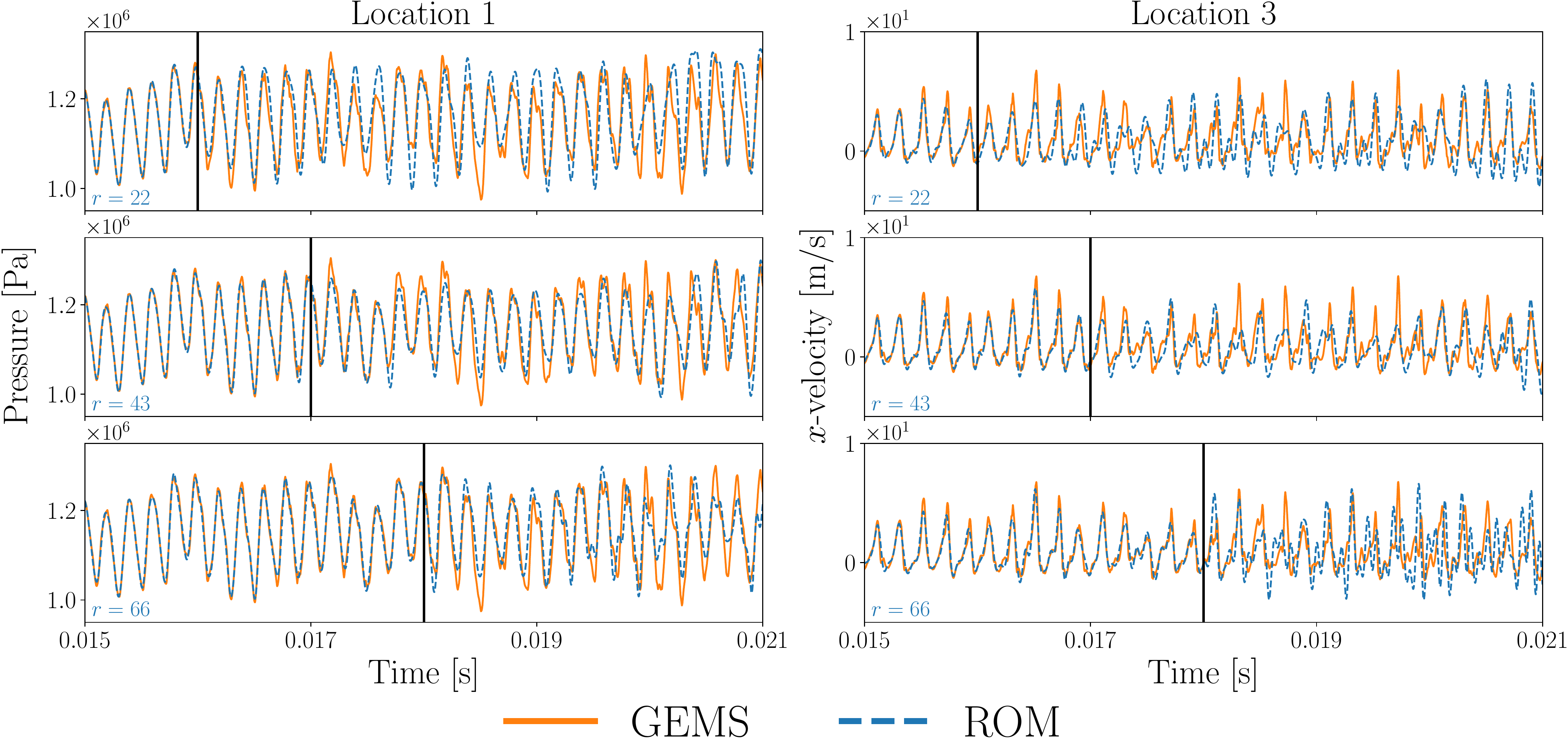}
    \caption{Traces of pressure and $x$-velocity through time at monitor locations $1$ and $3$ of Figure~\ref{fig:domain}, respectively, for $k = 10{,}000$ (top row), $k = 20{,}000$ (middle row), and $k = 30{,}000$ (bottom row) training snapshots. The vertical black lines separate the training and prediction periods. For each choice of $k$, the error over the training domain is low, but increasing $k$ has a significant impact on the behavior in the testing domain. Here $r$ is chosen in each case so that $\mathcal{E}_{r} > 0.985$.}
    \label{fig:time-traces}
\end{figure}

Figure~\ref{fig:time-traces} plots the GEMS and OpInf ROM results for pressure and $x$-velocity predictions over time at two of the monitor locations in Figure~\ref{fig:domain}.\footnote{See \url{https://github.com/Willcox-Research-Group/ROM-OpInf-Combustion-2D} for additional results.} While it can be misleading to assess accuracy based on predictions at a single spatial point, these plots reveal several representative insights.
First, each OpInf ROM faithfully reconstructs the training data but has some discrepancies in the prediction regime.
Second, the pressure and $x$-velocity frequencies are well captured throughout the time domain, but the amplitudes are sometimes less accurate in the prediction regime. The effects of the 5,000~Hz downstream pressure forcing are clearly visible in the pressure.
Third, we see the importance of the training data---as the amount of training data increases, the ROM predictions change significantly and generally (but not always) improve. This is yet another indication of the complexity of the dynamics we are aiming to approximate.

\begin{figure}
    \centering
    \includegraphics[width=.575\textwidth]{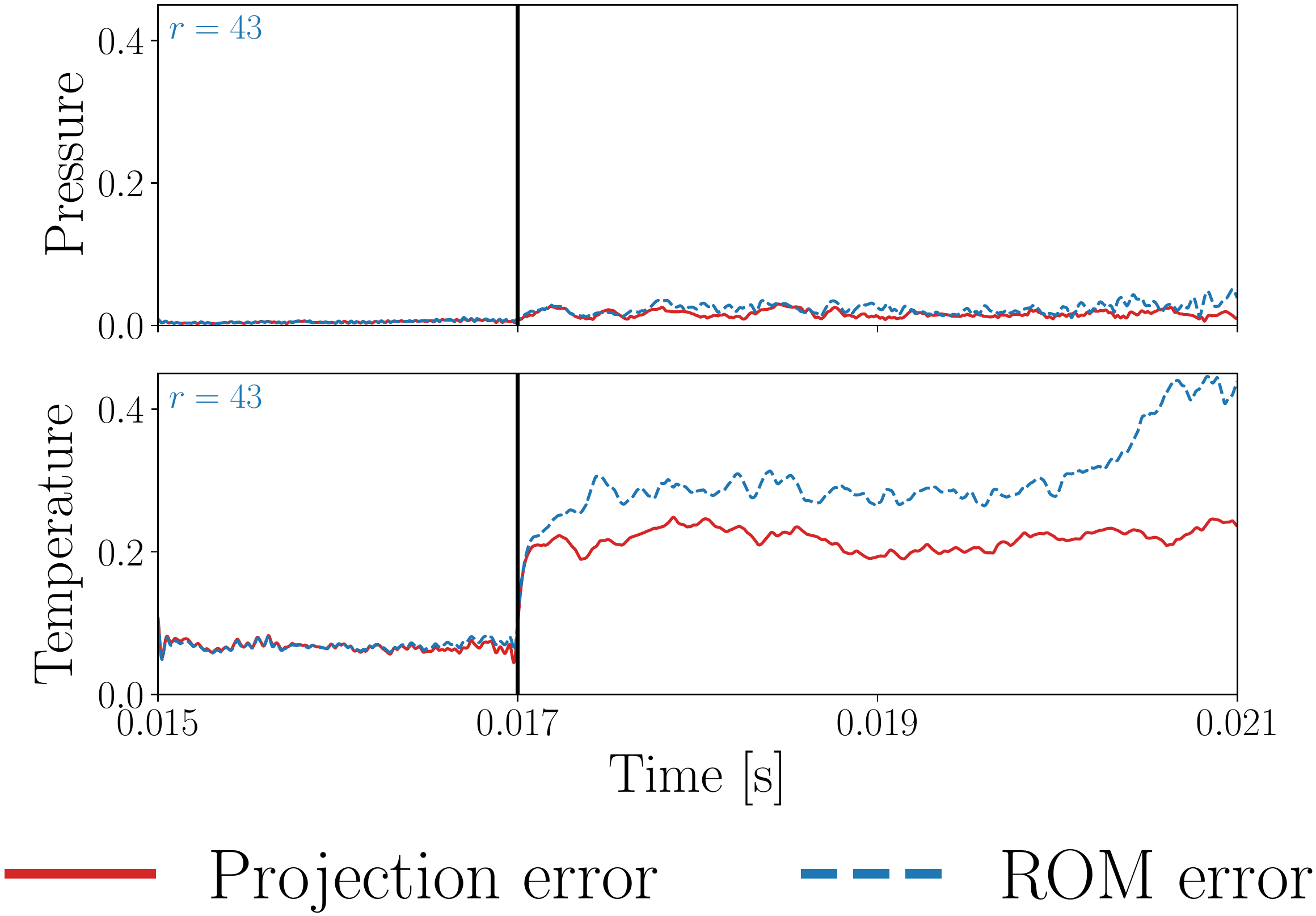}
    \hfill
    \includegraphics[width=.375\textwidth]{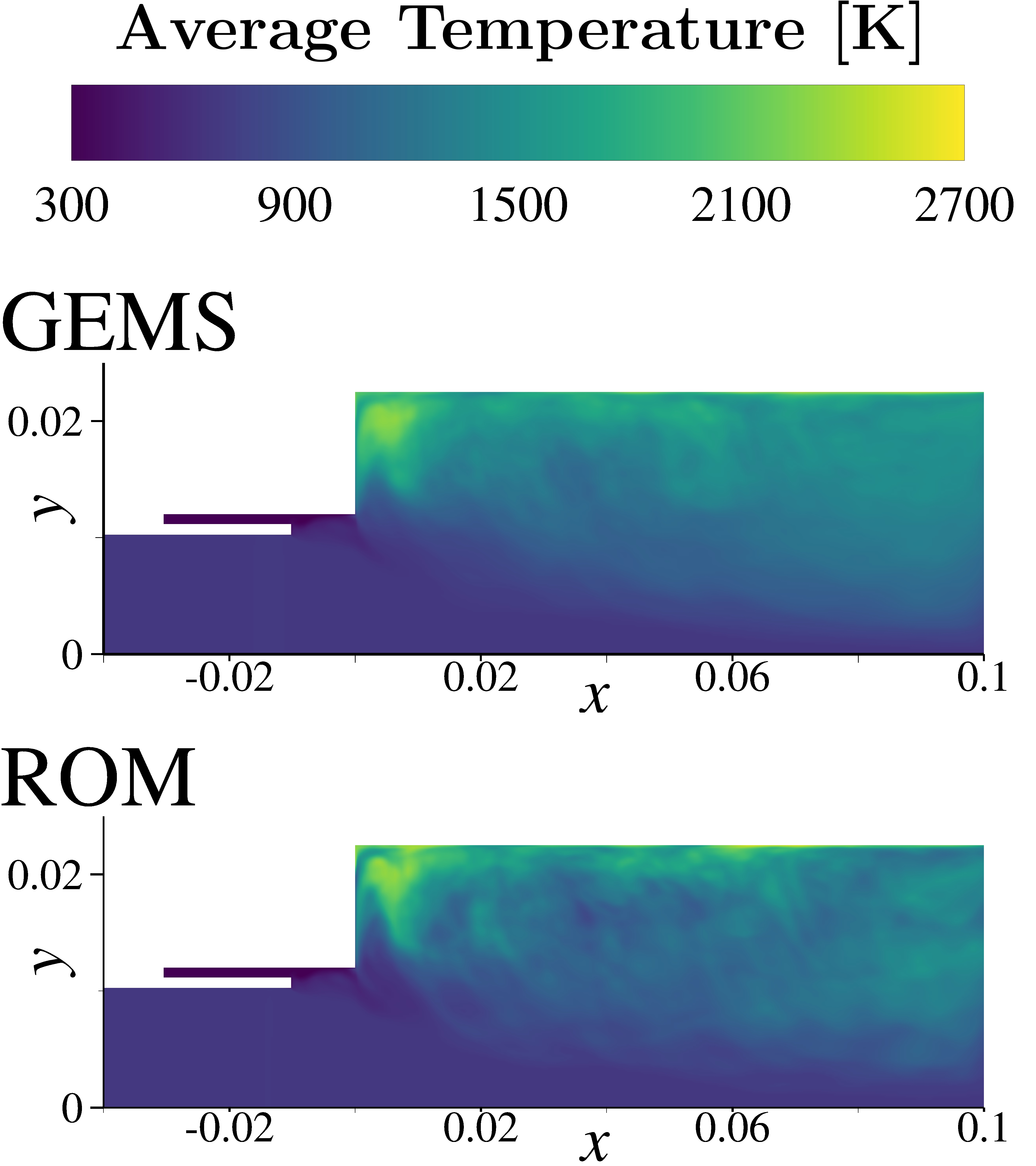}
    \caption{Spatially averaged relative errors in time for pressure and temperature (left) and temperature profiles for the GEMS dataset and an OpInf ROM averaged over $60{,}000$ time steps (right). The basis comprises the $r=43$ dominant singular vectors of $k = 20{,}000$ training snapshots.}
    \label{fig:projection-error}
\end{figure}

The effectiveness of the low-dimensional basis differs among the state variables.
Let $\mathbf{z}(t)\in\mathbb{R}^{8n_x}$ denote the GEMS simulation data at time $t$ with components $\begin{bmatrix}s_{1}(t) & \cdots & s_{n_x}(t)\end{bmatrix}^\top\in\mathbb{R}^{n_x}$ corresponding to a single state variable (e.g., pressure or temperature), and let $\mathcal{T}:\mathbb{R}^{8n_{x}}\to\mathbb{R}^{9n_{x}}$ be the preprocessing variable transformation and scaling with reverse operator $\mathcal{T}^\ast:\mathbb{R}^{9n_{x}}\to\mathbb{R}^{8n_{x}}$ (i.e., $\mathcal{T}^{\ast}(\mathcal{T}(\mathbf{z}(t))) = \mathbf{z}(t)$ for all $t$). Given the basis $\mathbf{V}\in\mathbb{R}^{9n_{x} \times r}$ for the scaled learning variables, the projection of the GEMS data onto the basis is
$
    \mathcal{T}^{\ast}\left(\mathbf{V}\mathbf{V}^\top
        \mathcal{T}\left(\mathbf{z}(t)\right)\right).
$
Let $\begin{bmatrix}s^{\textrm{proj}}_1(t) & \cdots & s^{\textrm{proj}}_{n_x}(t)\end{bmatrix}^\top\in\mathbb{R}^{n_x}$ denote the components of the projection corresponding to the state variable of interest.
Next, letting $\widetilde{\mathbf{q}}(t)$ denote the ROM state (the result of integrating Eq.~\eqref{eq:poly_ROM}), the reconstructed ROM prediction in the original state variables is
$
    \mathcal{T}^{\ast}\left(\mathbf{V}\widetilde{\mathbf{q}}(t)\right).
$
Let $\begin{bmatrix}s^{\textrm{pred}}_1(t) & \cdots & s^{\textrm{pred}}_{n_x}(t)\end{bmatrix}^\top\in\mathbb{R}^{n_x}$ be the components of this reconstruction corresponding to the state variable of interest.
We define the spatially averaged relative errors for the projection and the ROM predictions as
\begin{align*}
    s_{\textrm{projerr}}(t)
    &= \frac{1}{n_x}\sum_{i=1}^{n_x}\frac{|s^{\textrm{proj}}_{i}(t) - s_{i}(t)|}{|s_{i}(t)|},
    &
    s_{\textrm{prederr}}(t)
    &= \frac{1}{n_x}\sum_{i=1}^{n_x}\frac{|s^{\textrm{pred}}_{i}(t) - s_{i}(t)|}{|s_{i}(t)|},
\end{align*}
respectively.
Figure~\ref{fig:projection-error} plots these errors against time for pressure and temperature using a POD basis with $r = 43$ vectors computed from the first $k = 20{,}000$ training snapshots.
While both error measures for pressure remain low throughout the full time interval, both the projection errors and the ROM prediction errors for  temperature increase significantly at the end of the training regime. The temperature profile is influenced by both the advective flow dynamics and the local chemical reactions, which in combination lead to a highly nonlinear and multiscale behavior that is difficult to represent with the POD basis after the training period. However, while the ROM struggles to accurately predict the detailed temperature variations pointwise, it does adequately predict the general trends of temperature evolution beyond the training horizon. Figure~\ref{fig:projection-error} shows the time-averaged temperature profiles for the GEMS data and for an OpInf ROM, suggesting that the ROM captures the time-averaged behavior of the temperature dynamics.

\begin{figure} \centering
    \includegraphics[width=1\textwidth]{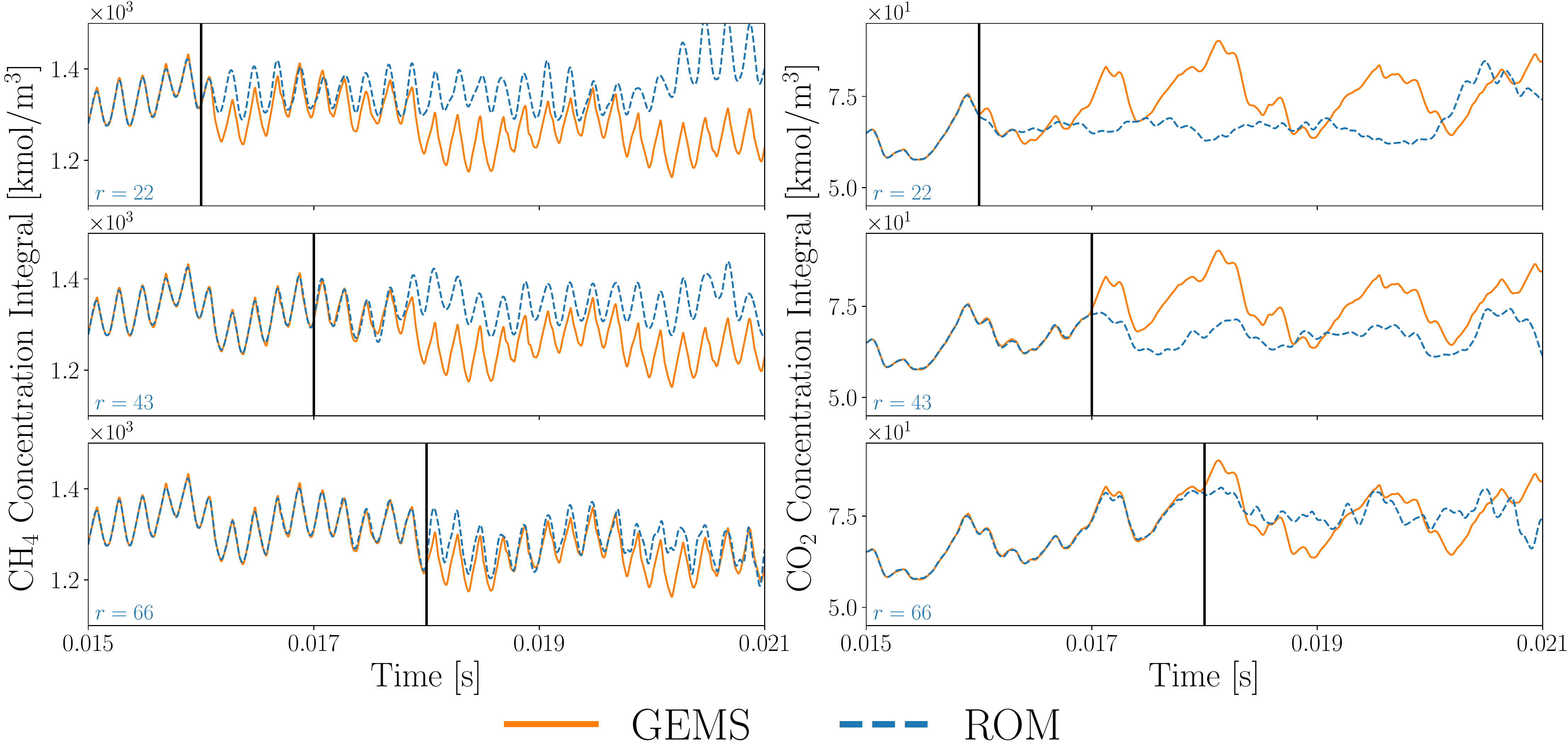}
    \caption{Integrated molar concentrations for CH$_4$ and CO$_2$, computed over the spatial domain for each point in time, for $k = 10{,}000$ (top row), $k = 20{,}000$ (middle row), and $k = 30{,}000$ (bottom row) training snapshots. These statistical features further highlight the effect of increasing $k$. As in Figure~\ref{fig:time-traces}, $r$ is chosen so that $\mathcal{E}_{r} > 0.985$.}
    \label{fig:feature-traces}
\end{figure}

Figure~\ref{fig:feature-traces} plots CH$_4$ and CO$_2$ concentrations integrated over the spatial domain. These measures give a more global sense of the ROM predictive accuracy and the predicted chemical reaction rate. In each case, the ROMs are able to accurately re-predict the training data and capture much of the overall system behavior in the prediction phase, with slightly more training error as the number of snapshots increases.
\subsection{Comparison to POD-DEIM}
\label{subsec:compare-poddeim}

We now compare regularized OpInf to a state-of-the-art nonlinear model reduction method that uses a least-squares Petrov-Galerkin POD projection coupled with the discrete empirical interpolation method (DEIM) \citep{CS2010deim}, as implemented for the same combustion problem in \cite{HDM2019poddeim-robustness,HXDM2018rocketrom-poddeim}. This POD-DEIM method is intrusive---it requires nonlinear residual evaluations of the GEMS code at sparse discrete interpolation points.
This also increases the computational cost of solving the POD-DEIM ROM in comparison to the OpInf ROM: integrating a POD-DEIM ROM with $r = 70$ for 6,000 time steps of size $\delta t = 10^{-6}$~s takes approximately $30$ minutes on two nodes, each with two Haswell CPUs processors at $2.60$~GHz and $20$ cores per node; for OpInf, using Python $3.6.9$ and a single CPU on an AMD EPYC 7,702 64-core processor at $3.3$~GHz with $2.1$~TB RAM, we solve Eq.~\eqref{eq:OpInf-reg} with $k = 20{,}000$ training snapshots and $r = 43$ POD modes in approximately $0.6$~s and integrate the resulting OpInf ROM for 60,000 time steps of size $\delta t = 10^{-7}$~s in approximately $0.4$~s. While these measurements are made on different hardware, and though the execution time for POD-DEIM can be improved with optimal load balancing, the difference in execution times (30~minutes versus 1~second) is representative and illustrates one of the advantages over POD-DEIM of the polynomial form employed in the OpInf approach.

\begin{figure}
\centering
    \includegraphics[width=1\textwidth]{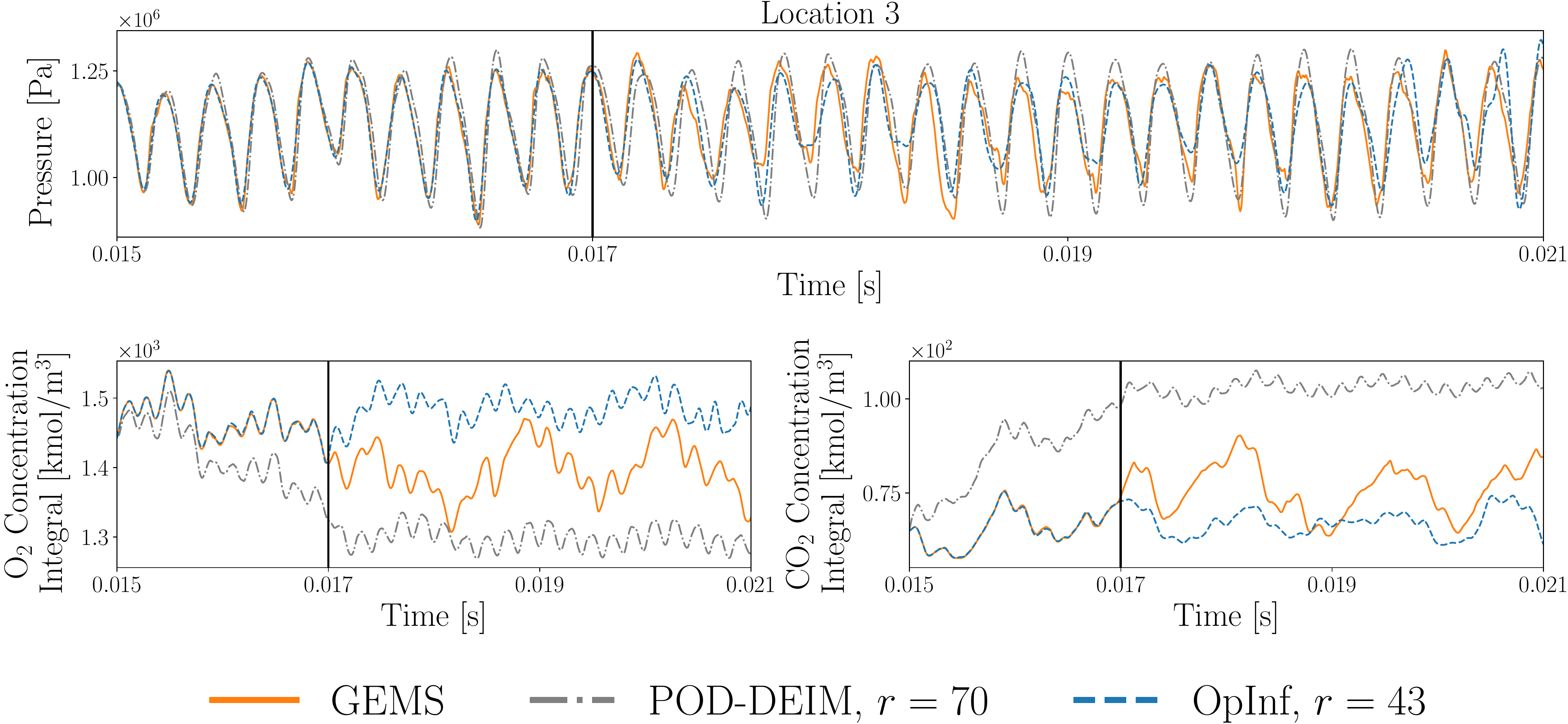}
\caption{Pressure trace at monitor location $3$ of Figure~\ref{fig:domain} (top) and spatially integrated O$_2$ and CO$_2$ molar concentrations (bottom), computed by GEMS, a POD-DEIM ROM, and an OpInf ROM. Both ROMs use $k = 20{,}000$ training snapshots with $r$ chosen so that $\mathcal{E}_{r} > 0.985$.}
    \label{fig:poddeim-traces}
\end{figure}

Figure~\ref{fig:poddeim-traces} compares select GEMS outputs to POD-DEIM and OpInf ROM outputs, with each ROM trained on $k = 20{,}000$ training snapshots. As before, the OpInf ROM dimension $r = 43$ is chosen such that $\mathcal{E}_{r} > 0.985$; the POD-DEIM ROM, which uses an entirely different basis than the OpInf approach, requires $r = 70$ vectors to achieve the same level of cumulative energy. Both approaches maintain appropriate pressure oscillation frequencies, and while neither model accurately predicts the global species concentration dynamics after the training period, the OpInf ROM reconstructs the training data more faithfully than the POD-DEIM ROM. \cite{HDM2019poddeim-robustness} show similar results for the same POD-DEIM model with 1~ms of training and 1~ms of prediction; here we are using 2~ms of training and 4~ms of prediction. Note from Figures \ref{fig:time-traces} and \ref{fig:feature-traces} that the OpInf ROMs achieve excellent prediction results for the 1~ms period following the training.

\begin{figure}
\centering
    \includegraphics[width=.95\textwidth]{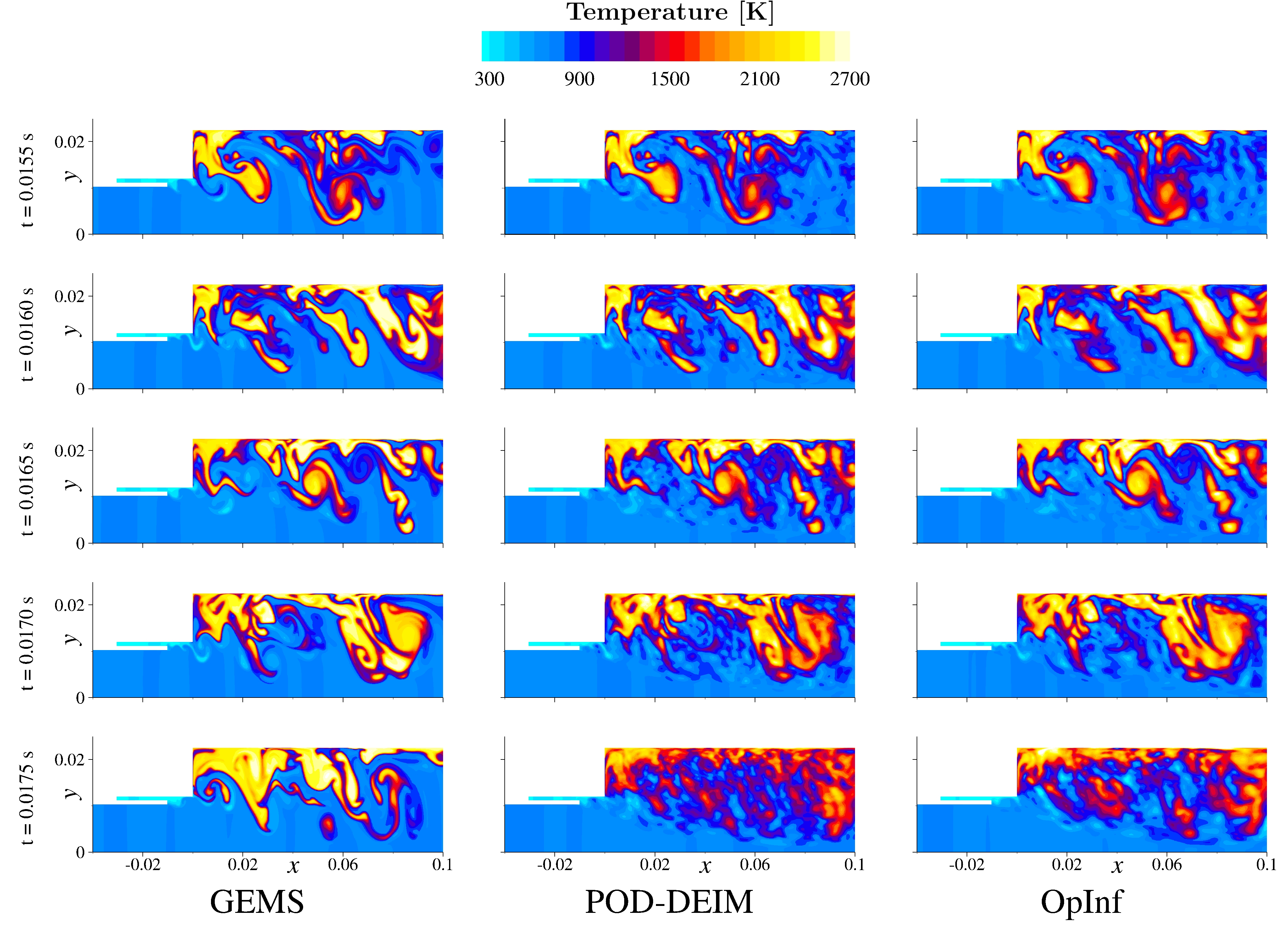}
\vspace*{-.5cm}
\caption{Temperature fields produced by GEMS (left column), POD-DEIM (middle column), and OpInf (right column), where each ROMs uses $k = 20{,}000$ training snapshots, with $r$ chosen so that $\mathcal{E}_{r} > 0.985$. Each row shows results for a given time, with an increment of $0.0005$~s between rows. The training period ends at $t = 0.0170$~s (fourth row); $t = 0.0175$~s (last row) is well into the prediction regime.}
\label{fig:temperature-fulldomain}
\end{figure}

\begin{figure}
\centering
    \includegraphics[width=.95\textwidth]{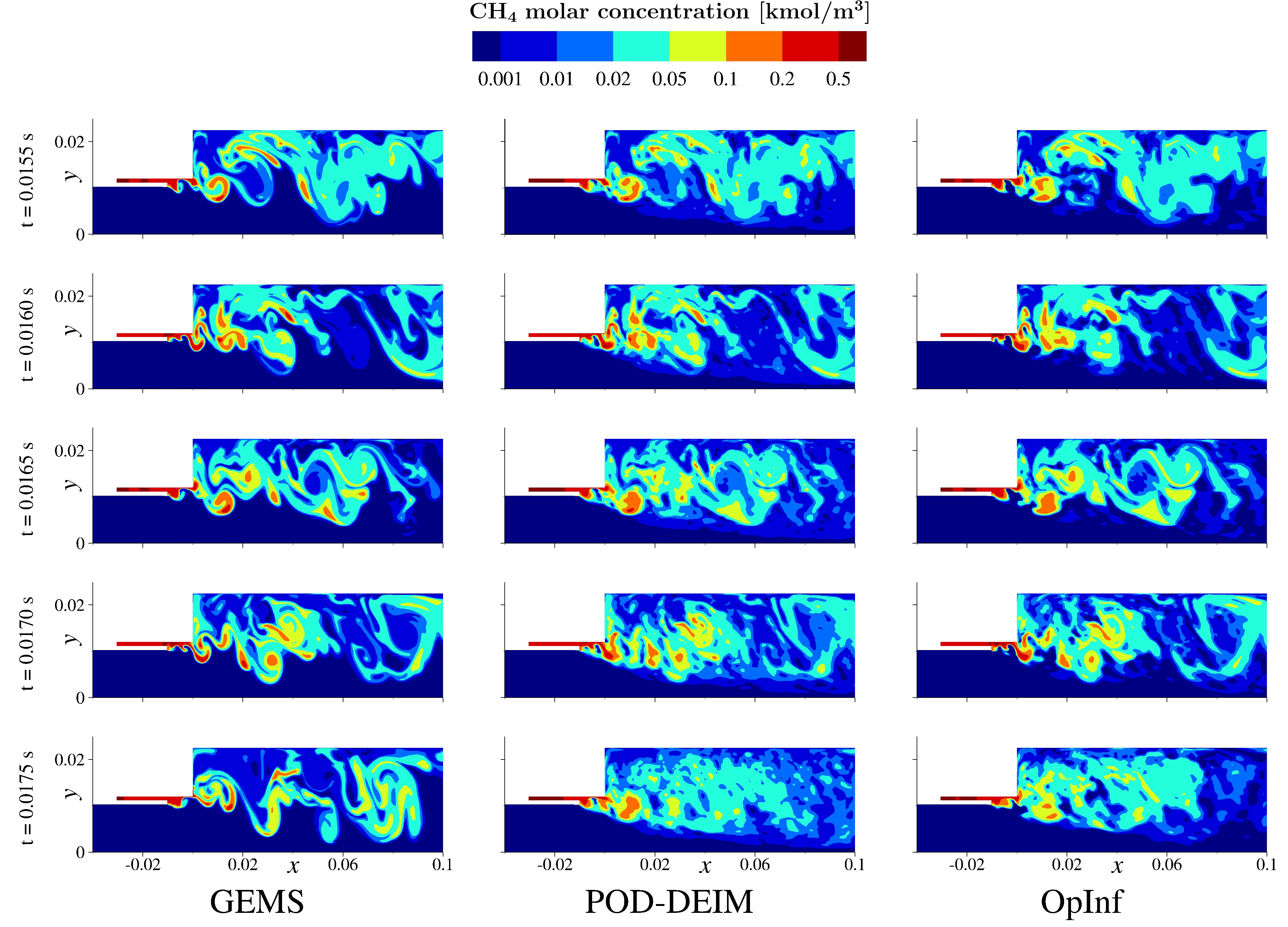}
\vspace*{-.5cm}
\caption{Molar concentrations of CH$_4$ produced by GEMS (left column), POD-DEIM (middle column), and OpInf (right column), where each ROMs uses $k = 20{,}000$ training snapshots, with $r$ chosen so that $\mathcal{E}_{r} > 0.985$. Each row shows results for a given time, with an increment of $0.0005$~s between rows. The training period ends at $t = 0.0170$~s (fourth row); $t = 0.0175$~s (last row) is well into the prediction regime.}
\label{fig:CH4-fulldomain}
\end{figure}

Figures~\ref{fig:temperature-fulldomain} and \ref{fig:CH4-fulldomain} show, respectively, full-domain results for the temperature and molar concentration of CH$_4$. The figures show the solution at time instants within the training regime, at the end of the training regime, and into the prediction regime. As with the point traces shown earlier, we see that the ROMs have impressive accuracy over the training region, but lose accuracy as they attempt to predict dynamics beyond the training horizon. However, many of the coherent features are reasonably predicted, especially the recirculation zone dynamics near the dump plane ($x = 0$ in Figure~\ref{fig:domain}) shown in the temperature fields. Significantly, both ROMs maintain appropriate temperature ranges throughout the prediction phase. The POD-DEIM ROM explicitly enforces such limits by reconstructing the full solution at each time step, constraining the temperature to a desired range, and projecting the result back to the reduced space (see Section IV.G of \cite{HDM2019poddeim-robustness}); in contrast, the OpInf ROM selects a regularization that results in bounded behavior due to the criteria $|\hat{q}_{i}(t)| \le B$ (see Eq.~\eqref{eq:state-bounding}). In other words, POD-DEIM limits the temperature in the online phase, while OpInf builds a similar constraint into the offline phase.

\newpage
\section{Conclusions} \label{sec:conclusions}

The presented scientific machine learning approach is broadly applicable to problems where the governing equations are known but access to the high-fidelity simulation code is limited.
The approach is computationally as accessible as black-box surrogate modeling while achieving the accuracy of intrusive projection-based model reduction.
While the conclusions drawn from the numerical studies apply to the single-injector combustion example, they are relevant and likely apply to other problems.
First, the quality and quantity of the training data are critical to the success of the method.
Second, regularization is essential to avoid overfitting.
Third, achieving a low error over the training regime is not necessarily indicative of a reduced model with good predictive capability.
This emphasizes the importance of the training data.
Fourth, physical quantities that exhibit large-scale coherent structures (e.g., pressure) are more accurately predicted by a reduced-order model than quantities that exhibit multiscale behavior (e.g., temperature, species concentrations).
Fifth, a significant advantage of the data-driven learning aspects of the approach is that the reduced model may be derived in any variables.
This includes the possibility to include redundancy in the learning variables (e.g., to include both pressure and temperature).
Overall, this paper illustrates the power and effectiveness of learning from data through the lens of physics-based models as a physics-grounded alternative to black-box machine learning.

\section*{Acknowledgements}

This work has been supported in part by the Air Force Center of Excellence on Multi-Fidelity Modeling of Rocket Combustor Dynamics under award FA9550-17-1-0195, and the US Department of Energy AEOLUS MMICC center under award DE-SC0019303.

\bibliographystyle{tfcse}
\bibliography{references}

\begin{thebibliography}{44}
\providecommand{\natexlab}[1]{#1}
\providecommand{\url}[1]{\normalfont{#1}}
\providecommand{\urlprefix}{Available from: }

\bibitem[Antoulas(2005)]{antoulas2005approximation}
Antoulas~AC. 2005. Approximation of large-scale dynamical systems.
  Philadelphia, PA: SIAM.

\bibitem[Astrid et~al.(2008)]{Astrid2008}
Astrid~P, Weiland~S, Willcox~K, Backx~T. 2008. Missing point estimation in
  models described by proper orthogonal decomposition. IEEE Transactions on
  Automatic Control. 53(10):2237--2251.

\bibitem[Bai(2002)]{Bai02}
Bai~Z. 2002. Krylov subspace techniques for reduced-order modeling of
  large-scale dynamical systems. Applied Numerical Mathematics. 43(1--2):9--44.

\bibitem[Baker et~al.(2019)]{baker2019workshop}
Baker~N, Alexander~F, Bremer~T, Hagberg~A, Kevrekidis~Y, Najm~H, Parashar~M,
  Patra~A, Sethian~J, Wild~S, et~al. 2019. Workshop report on basic research
  needs for scientific machine learning: {C}ore technologies for artificial
  intelligence. USDOE Office of Science (SC), Washington, DC (United States).
  Report {N}o:. {https://www.osti.gov/servlets/purl/1478744}.

\bibitem[Barrault et~al.(2004)]{BMNP2004eim}
Barrault~M, Maday~Y, Nguyen~NC, Patera~AT. 2004. An `empirical interpolation'
  method: application to efficient reduced-basis discretization of partial
  differential equations. Comptes Rendus Mathematique. 339(9):667--672.

\bibitem[Benner et~al.(2015)]{BGW2015pmorSurvery}
Benner~P, Gugercin~S, Willcox~K. 2015. A survey of projection-based model
  reduction methods for parametric dynamical systems. SIAM Review.
  57(4):483--531.

\bibitem[Berkooz et~al.(1993)]{berkooz}
Berkooz~G, Holmes~P, Lumley~J. 1993. The proper orthogonal decomposition in the
  analysis of turbulent flows. Annual Review of Fluid Mechanics. 25:539--575.

\bibitem[Brunton et~al.(2016)]{BPK2016sindy}
Brunton~SL, Proctor~JL, Kutz~JN. 2016. Discovering governing equations from
  data by sparse identification of nonlinear dynamical systems. Proceedings of
  the National Academy of Sciences. 113(15):3932--3937.

\bibitem[Carlberg et~al.(2018)]{CCS2018conservativeROM}
Carlberg~K, Choi~Y, Sargsyan~S. 2018. Conservative model reduction for
  finite-volume models. Journal of Computational Physics. 371:280--314.

\bibitem[Carlberg et~al.(2013)]{Carlberg2013}
Carlberg~K, Farhat~C, Cortial~J, Amsallem~D. 2013. The {GNAT} method for
  nonlinear model reduction: {E}ffective implementation and application to
  computational fluid dynamics and turbulent flows. Journal of Computational
  Physics. 242:623--647.

\bibitem[Chaturantabut and Sorensen(2010)]{CS2010deim}
Chaturantabut~S, Sorensen~DC. 2010. Nonlinear model reduction via discrete
  empirical interpolation. SIAM Journal on Scientific Computing.
  32(5):2737--2764.

\bibitem[Cui et~al.(2011)]{CFOS2011geothermalbayes}
Cui~T, Fox~C, O'{S}ullivan~MJ. 2011. {B}ayesian calibration of a large-scale
  geothermal reservoir model by a new adaptive delayed acceptance {M}etropolis
  {H}astings algorithm. Water Resources Research. 47(10).

\bibitem[Deane et~al.(1991)]{deane1991low}
Deane~A, Kevrekidis~I, Karniadakis~GE, Orszag~S. 1991. Low-dimensional models
  for complex geometry flows: {A}pplication to grooved channels and circular
  cylinders. Physics of Fluids A: Fluid Dynamics. 3(10):2337--2354.

\bibitem[Demmel(1997)]{demmel1997nla}
Demmel~JW. 1997. Applied numerical linear algebra. Philadelphia, PA: SIAM.

\bibitem[Dowell and Hall(2001)]{dowell2001modeling}
Dowell~EH, Hall~KC. 2001. Modeling of fluid-structure interaction. Annual
  Review of Fluid Mechanics. 33(1):445--490.

\bibitem[Freund(2003)]{Fre03}
Freund~R. 2003. Model reduction methods based on {K}rylov subspaces. Acta
  Numerica. 12:267--319.

\bibitem[Gatski and Glauser(1992)]{gatski1992pod-turbulence}
Gatski~T, Glauser~M. 1992. Proper orthogonal decomposition based turbulence
  modeling. In: Instability, transition, and turbulence. Springer; p. 498--510.

\bibitem[Grepl and Patera(2005)]{Grepl05}
Grepl~M, Patera~A. 2005. {\em A posteriori} error bounds for reduced-basis
  approximations of parametrized parabolic partial differential equations.
  ESAIM-Mathematical Modelling and Numerical Analysis (M2AN). 39(1):157--181.

\bibitem[Halko et~al.(2011)]{HMPT2011rNLA}
Halko~N, Martinsson~PG, Tropp~JA. 2011. Finding structure with randomness:
  {P}robabilistic algorithms for constructing approximate matrix
  decompositions. SIAM Review. 53(2):217--288.

\bibitem[Harvazinski et~al.(2015)]{HHSFAMT2015gems}
Harvazinski~ME, Huang~C, Sankaran~V, Feldman~TW, Anderson~WE, Merkle~CL,
  Talley~DG. 2015. Coupling between hydrodynamics, acoustics, and heat release
  in a self-excited unstable combustor. Physics of Fluids. 27(4):045102.

\bibitem[Huang et~al.(2019)]{HDM2019poddeim-robustness}
Huang~C, Duraisamy~K, Merkle~CL. 2019. Investigations and improvement of
  robustness of reduced-order models of reacting flow. AIAA Journal.
  57(12):5377--5389.

\bibitem[Huang et~al.(2018)]{HXDM2018rocketrom-poddeim}
Huang~C, Xu~J, Duraisamy~K, Merkle~C. 2018. Exploration of reduced-order models
  for rocket combustion applications. In: 2018 AIAA Aerospace Sciences Meeting;
  Orlando, FL. Paper AIAA-2018-1183.

\bibitem[Kalashnikova et~al.(2014)]{KvBWAB2014eigreassign}
Kalashnikova~I, van Bloemen~Waanders~B, Arunajatesan~S, Barone~M. 2014.
  Stabilization of projection-based reduced order models for linear
  time-invariant systems via optimization-based eigenvalue reassignment.
  Computer Methods in Applied Mechanics and Engineering. 272:251--270.

\bibitem[Kramer and Willcox(2019)]{Kramer-AIAA2019}
Kramer~B, Willcox~K. 2019. Nonlinear model order reduction via lifting
  transformations and proper orthogonal decomposition. AIAA Journal.
  57(6):2297--2307.

\bibitem[Lumley(1967)]{lumley}
Lumley~J. 1967. The structures of inhomogeneous turbulent flow. Atmospheric
  Turbulence and Radio Wave Propagation:166--178.

\bibitem[Nelder and Mead(1965)]{nelder1965simplex}
Nelder~JA, Mead~R. 1965. A simplex method for function minimization. The
  Computer Journal. 7(4):308--313.

\bibitem[Nordsletten et~al.(2011)]{nordsletten2011fluid}
Nordsletten~D, McCormick~M, Kilner~P, Hunter~P, Kay~D, Smith~N. 2011.
  Fluid--solid coupling for the investigation of diastolic and systolic human
  left ventricular function. International Journal for Numerical Methods in
  Biomedical Engineering. 27(7):1017--1039.

\bibitem[O'{S}ullivan et~al.(2001)]{sullivan2001state}
O'{S}ullivan~M, Pruess~K, Lippmann~M. 2001. State of the art of geothermal
  reservoir simulation. Geothermics. 30:395--429.

\bibitem[Pedregosa et~al.(2011)]{sklearn-api}
Pedregosa~F, Varoquaux~G, Gramfort~A, Michel~V, Thirion~B, Grisel~O, Blondel~M,
  Prettenhofer~P, Weiss~R, Dubourg~V, et~al. 2011. Scikit-learn: Machine
  learning in {P}ython. Journal of Machine Learning Research. 12:2825--2830.

\bibitem[Peherstorfer(2020)]{Peherstorfer2019reprojection}
Peherstorfer~B. 2020. Sampling low-dimensional {M}arkovian dynamics for
  pre-asymptotically recovering reduced models from data with operator
  inference. SIAM Journal on Scientific Computing. 42(5):A3489--A3515.

\bibitem[Peherstorfer and Willcox(2016)]{PW2016operatorInference}
Peherstorfer~B, Willcox~K. 2016. Data-driven operator inference for
  nonintrusive projection-based model reduction. Computer Methods in Applied
  Mechanics and Engineering. 306:196--215.

\bibitem[Qian et~al.(2020)]{QKPW2020liftAndLearn}
Qian~E, Kramer~B, Peherstorfer~B, Willcox~K. 2020. Lift \& {L}earn:
  {P}hysics-informed machine learning for large-scale nonlinear dynamical
  systems. Physica {D}: {N}onlinear {P}henomena. 406:132401.

\bibitem[Rezaian and Wei(2020)]{rezaian2020hybrid}
Rezaian~E, Wei~M. 2020. A hybrid stabilization approach for reduced-order
  models of compressible flows with shock-vortex interaction. International
  Journal for Numerical Methods in Engineering. 121(8):1629--1646.

\bibitem[Rozza et~al.(2008)]{RozHP08}
Rozza~G, Huynh~D, Patera~A. 2008. Reduced basis approximation and a posteriori
  error estimation for affinely parametrized elliptic coercive partial
  differential equations: application to transport and continuum mechanics.
  Archives of Computational Methods in Engineering. 15(3):229--275.
  \urlprefix\url{http://dx.doi.org/10.1007/s11831-008-9019-9}.

\bibitem[Sirovich(1987)]{sirovich1987turbulence}
Sirovich~L. 1987. Turbulence and the dynamics of coherent structures. {I}.
  {C}oherent structures. Quarterly of Applied Mathematics. 45(3):561--571.

\bibitem[Spalart and Venkatakrishnan(2016)]{spalart2016role}
Spalart~P, Venkatakrishnan~V. 2016. On the role and challenges of cfd in the
  aerospace industry. The Aeronautical Journal. 120(1223):209.

\bibitem[Swischuk et~al.(2020)]{SKHW2020romCombustion}
Swischuk~R, Kramer~B, Huang~C, Willcox~K. 2020. Learning physics-based
  reduced-order models for a single-injector combustion process. AIAA Journal.
  58(6):2658--2672.

\bibitem[Tikhonov and Arsenin(1977)]{Tikhonov1977regularization}
Tikhonov~AN, Arsenin~VY. 1977. Solutions of ill-posed inverse problems. New
  York, NY: Wiley.

\bibitem[Veroy and Patera(2005)]{Veroy05}
Veroy~K, Patera~A. 2005. Certified real-time solution of the parametrized
  steady incompressible {N}avier-{S}tokes equations: {R}igorous reduced-basis
  {\em a posteriori} error bounds. International Journal for Numerical Methods
  in Fluids. 47:773--788.

\bibitem[Veroy et~al.(2003)]{Veroy03}
Veroy~K, Prud'homme~C, Rovas~D, Patera~A. 2003. A posteriori error bounds for
  reduced-basis approximation of parametrized noncoercive and nonlinear
  elliptic partial differential equations. In: Proceedings of the 16th AIAA
  Computational Fluid Dynamics Conference; Orlando, FL. Paper AIAA-2003-3847.

\bibitem[{Virtanen} et~al.(2020)]{2020SciPy-NMeth}
{Virtanen}~P, {Gommers}~R, {Oliphant}~TE, {Haberland}~M, {Reddy}~T,
  {Cournapeau}~D, {Burovski}~E, {Peterson}~P, {Weckesser}~W, {Bright}~J, et~al.
  2020. {Sci{P}y 1.0: {F}undamental Algorithms for Scientific Computing in
  {P}ython}. Nature Methods. 17:261--272.

\bibitem[Walt et~al.(2011)]{WCV2011NumPy}
Walt~Svd, Colbert~SC, Varoquaux~G. 2011. The {N}um{P}y array: {A} structure for
  efficient numerical computation. Computing in Science \& Engineering.
  13(2):22--30.

\bibitem[Westbrook and Dryer(1981)]{WD1981oxidation}
Westbrook~CK, Dryer~FL. 1981. Simplified reaction mechanisms for the oxidation
  of hydrocarbon fuels in flames. Combustion Science and Technology.
  27(1-2):31--43.

\bibitem[Yin et~al.(2010)]{yin2010simulation}
Yin~Y, Choi~J, Hoffman~EA, Tawhai~MH, Lin~CL. 2010. Simulation of pulmonary air
  flow with a subject-specific boundary condition. Journal of Biomechanics.
  43(11):2159--2163.

\end{thebibliography}

\end{document}